\documentclass[english,11pt,oneside]{article}

\pdfoutput=1

\usepackage{amsmath}
\usepackage{amsfonts}
\usepackage{amssymb}
\usepackage{graphicx, rotating}
\usepackage{epstopdf}
\usepackage{epsfig}
\usepackage{latexsym}
\usepackage{multirow}
\usepackage{color}
\usepackage[dvipsnames]{xcolor}
\usepackage{slashed}
\usepackage[top=2.4 cm, bottom=2.4 cm, left=2.5 cm, right=2.5 cm]{geometry}
\usepackage{amstext}
\usepackage{array}  
\usepackage{hyperref}
\hypersetup{
colorlinks = true,
 linkcolor  = red, %black
 linktocpage=true,
 citecolor={blue}
}
\usepackage{bbold}

\usepackage[utf8]{inputenc}
\usepackage{bm}
\usepackage{leftidx}
\usepackage[numbers,sort&compress]{natbib}

\usepackage{makecell}

\usepackage{authblk}
\usepackage{colortbl}
\usepackage{xcolor}

\usepackage{float}

\newcommand{\be}{\begin{equation}}
\newcommand{\ee}{\end{equation}}
\newcommand{\ba}{\begin{array}}
\newcommand{\ea}{\end{array}}
\newcommand{\into}{\ensuremath{\,\rightarrow\,}}
\newcommand{\eq}[1]{Eq.~(\ref{#1})}

\newcommand{\SO}[1]{\ensuremath{\mathrm{SO}(#1)}}
\newcommand{\SU}[1]{\ensuremath{\mathrm{SU}(#1)}}
\newcommand{\U}[1]{\ensuremath{\mathrm{U}(#1)}}
\newcommand{\tr}{\operatorname{tr}}
\newcommand{\vev}[1]{\langle #1 \rangle}
\definecolor{LightGrey}{gray}{0.96}
\definecolor{Grey}{gray}{0.94}
\newcommand{\diag}{\operatorname{diag}}

\begin{document}

\title{\bf Accidentally light scalars from large representations}
\renewcommand\footnotemark{}

\author[1]{Felix Br\"ummer}
\author[1]{Giacomo Ferrante}
\author[2]{Michele Frigerio}
\author[3,4]{Thomas Hambye}
\affil[1]{ \emph{Laboratoire Univers et Particules de Montpellier (LUPM), University of Montpellier and CNRS, Montpellier, France}}
\affil[2]{ \emph{Laboratoire Charles Coulomb (L2C), University of Montpellier and CNRS, Montpellier, France}}
\affil[3]{ \emph{Service de Physique Théorique, Université Libre de Bruxelles, Boulevard du Triomphe, CP225, 1050 Brussels, Belgium}} 
\affil[4]{ \emph{Theoretical Physics Department, CERN, 1211 Geneva 23, Switzerland}}\vspace{0.2cm}
\date{\today}
\maketitle
\begin{abstract}
 \noindent 
In models with spontaneous symmetry breaking by scalar fields in large group representations, we observe that some of the scalar masses can be loop-suppressed with respect to the naive expectation from symmetry selection rules. We present minimal models -- the $\SU{2}$ five-plet and $\SU{3}$ ten-plet -- with such accidentally light scalars, featuring compact tree-level flat directions lifted by radiative corrections. We sketch some potential applications, from stable relics and slow roll in cosmology, to hierarchy and fine-tuning problems in particle physics. 
\end{abstract}

%\tableofcontents

%%%%%%%%%%%%%%%%%%%%%%%%%%%%%%%%%%%%%%%%%%%%%%%%
\section{Introduction}
%%%%%%%%%%%%%%%%%%%%%%%%%%%%%%%%%%%%%%%%%%%%%%%%

In this work we consider perturbative, renormalisable models of scalar fields in four-dimensional quantum field theory. When all operators allowed by Lorentz invariance and the internal symmetries of the model are included in the Lagrangian with generic coefficients, all scalar fields are massive, with the only exception of Nambu-Goldstone bosons (NGBs). By Goldstone's theorem, in models with a continuous global symmetry group $G$ spontaneously broken to a subgroup $H$, the NGBs, forming the coset $G/H$, are exactly massless to all orders.
As a general rule, the mass of non-Goldstone scalar fields arises at the tree level from the scalar potential. One well-known exception to this rule are pseudo-NGBs, which appear when the symmetry $F$ of the scalar potential is larger than the symmetry $G$ which defines the model as a whole \cite{Weinberg:1972fn}.
This requires additional fields and interactions, such as gauge or Yukawa couplings, which explicitly break $F$ and induce pseudo-NGB masses via loops.

It is less well known that there exist models with tree-level massless scalars which are neither NGBs nor pseudo-NGBs in the above sense. In these models, the most general renormalisable potential compatible with the symmetry $G$ is {\it not} invariant under an enhanced continuous symmetry larger than $G$. Still, some non-NGB scalar fields remain massless at the tree level. To distinguish these accidentally tree-level massless fields from pseudo-NGBs as defined above, we will call them ``accidents'' for short. Accidents were encountered e.g.~in pre-QCD attempts to build renormalisable models of mesons \cite{Bars:1973vp}, and their nature was clearly emphasised in an early precursor \cite{Georgi:1975tz} of the little-Higgs idea (reviewed in \cite{Schmaltz:2005ky}).\footnote{One may prefer to extend the definition of pseudo-NGB to any scalar which is massless at the leading order in the loop expansion. In this case accidents can be dubbed pseudo-NGBs too, as done e.g.~in Ref.~\cite{Georgi:1975tz}. We retain the name ``accident'', in order to distinguish this peculiar class of light particles. In most of the more recent little-Higgs constructions \cite{Schmaltz:2005ky}, the little Higgs is a conventional pseudo-NGB rather than an accident.}
Accidents also appear in O'Raifeartaigh-like models of spontaneous supersymmetry (SUSY) breaking \cite{Capper:1976mf}, where they have been dubbed ``pseudo-moduli'' and studied in greater detail more recently \cite{Ray:2006wk,Intriligator:2008fe}; in this case the scalar potential is  additionally constrained by SUSY.

In this paper we present some models with accidentally light scalars
which are in a sense the most minimal ones. We focus on two examples, with symmetry groups $G=\SU{2}\times\U{1}$ and $G=\SU{3}\times\U{1}$; there are no additional discrete symmetries imposed; the field content is a single scalar multiplet in an irreducible representation. This is to be contrasted with the examples in the literature, which to our knowledge tend to rely on more complicated continuous symmetries (typically multiple copies of the same group), feature additional ad-hoc discrete symmetries, and  contain several scalar fields in various representations. The price to pay to avoid these complications is to take the single scalar field in a large representation of $G$.

The simplest possibilities are the five-plet in the $\SU{2}$ case, 
analysed in section \ref{sec:su2}, and the ten-plet in the $\SU{3}$ case, analysed in section \ref{sec:su3}. There are few known results on spontaneous symmetry breaking by large representations, see e.g.~\cite{Jetzer:1983ij}, and these do not cover our cases of interest, which further motivates our analysis.

The geometry of field space is non-trivial in our models, due to the large field representation involved. There is a compact manifold $M'$ of degenerate tree-level vacua. The continuous symmetry group $G$ is completely broken at a generic point on $M'$. When $G$ is gauged, all points on a single $G$-orbit are identified, but the resulting tree-level vacuum manifold $M$ of physically inequivalent points does not reduce to a single point. Instead, $M$ is parameterised by one or more non-Goldstone flat directions in field space: these correspond to tree-level massless ``accident'' fields. Both the scalar and the vector mass spectrum change when moving along $M$. At special points in $M$, the vacuum symmetry is enhanced (i.e.~a subgroup $H$ of $G$ is unbroken), and additional accidents appear.

In section \ref{sec:apps} we will briefly discuss a few potential phenomenological applications of accidents in cosmology (dark matter, slow roll) and in particle physics (Higgs, doublet-triplet splitting). A detailed analysis of the phenomenology is left for future work.

%%%%%%%%%%%%%%%%%%%%%%%%%%%%%%%%%%%%%%%%%%%%%%%%
\section{The simplest model with accidents: an SU(2) five-plet}\label{sec:su2}
%%%%%%%%%%%%%%%%%%%%%%%%%%%%%%%%%%%%%%%%%%%%%%%%

\subsection{Potential and tree-level spectrum}\label{potT}

Our first example of a model with accidentally light scalars is for $G=\SU{2}\times\U{1}$
with a complex scalar in the five-dimensional representation of $\SU{2}$ and with unit $\U{1}$ charge, $\phi\sim {\bf 5}_1$.
The most general $G$-invariant renormalisable potential can be written as
\be
 V=-\mu^2\,S+\frac{1}{2}\left[\lambda \,S^2+ \kappa
 \left(S^2-|S'|^2\right)+ \delta\, A^a A^a\right]\,.
\label{V5}\ee
Here $S$, $S'$ and $A^a$ are bilinears transforming in the singlet and adjoint representation of $\SU{2}$,
\be
S=\phi^\dag\phi\,,\qquad S'=\phi^T \phi\,, \qquad A^a=\phi^\dag T^a\phi\quad(a=1,2,3)\,,
\label{bi5}\ee
where $T^a$ are the $\SU{2}$ five-plet generators, which may be chosen imaginary and antisymmetric and satisfy $\tr\left(T^a T^b \right) = 10\,\delta^{ab}$.
It can be checked that the $\SO{10}$ global symmetry of the free theory 
is broken explicitly by $V$ to $\SU{2}\times\U{1}$: there is no larger continuous accidental symmetry. 

Since our aim is to study spontaneous symmetry breaking, we take $\mu^2>0$.
We further take $\lambda$, $\kappa$ and $\delta$ to be positive. While $\lambda$ must be positive for the potential to be bounded from below, negative values for $\kappa$ and $\delta$ are possible, but not of interest here. With our choice, each of the three terms in the quartic potential is positive definite, and
\begin{itemize}
 \item $S$ is non-zero if and only if at least one vacuum expectation value (VEV) is non-vanishing;
 \item $S^2-|S'|^2$ vanishes if and only if $\phi$ and $\phi^*$ are aligned in field space, i.e.~$\phi=c\hat\varphi$ for some complex number $c$ and real unit vector $\hat\varphi$;
 \item if $\phi$ and $\phi^*$ are aligned, then $A^a=0$ by antisymmetry; this corresponds to the fact that the adjoint ${\bf 3}$ is contained in the antisymmetric part of the product ${\bf 5}\otimes{\bf 5}$.
\end{itemize}
A minimum is therefore found by choosing $\vev\phi$ such that $\phi$ and $\phi^*$ are aligned and $\vev{S}=\mu^2/\lambda$: 
\be
 \vev{\phi_j}=\frac{v_j}{\sqrt{2}}\,e^{i\theta}\quad(j=1\ldots 5)\,,\qquad  v_j\in\mathbb{R}\,,\qquad\theta\in[0,\,2\pi)\,,\qquad
v^2\equiv v_j v_j
 =\frac{2\mu^2}{\lambda}\,.
\label{minV}\ee
The remarkable feature of this potential is the existence of one flat direction which is \emph{not} associated to a NGB, corresponding to one accidentally massless field. The vacuum manifold is indeed five-dimensional: while the overall scale $v$ of the VEV is fixed by the minimisation condition, one is free to rotate the five components $v_j$, and to choose the angle $\theta$. Changes in $\theta$ correspond to the $\U{1}$ Goldstone direction. Of the four directions in $\SO{5}/\SO{4}\simeq S^4$ corresponding to different choices for the orientation of the VEV,\footnote{More precisely one should divide by a $\mathbb{Z}_2$ since, for any $v$, the points $(v,\theta)$ and $(-v,\theta+\pi)$ are identified.} three are associated to the $\SU{2}$ Goldstone directions, but the fourth one does not correspond to any symmetry generator. When gauging $\SU{2}\times\U{1}$, all points on the Goldstone manifold are identified, but there remains a flat direction of degenerate, physically inequivalent vacua.

\begin{figure}[ht]
\begin{center}
\includegraphics[width=.42\textwidth]{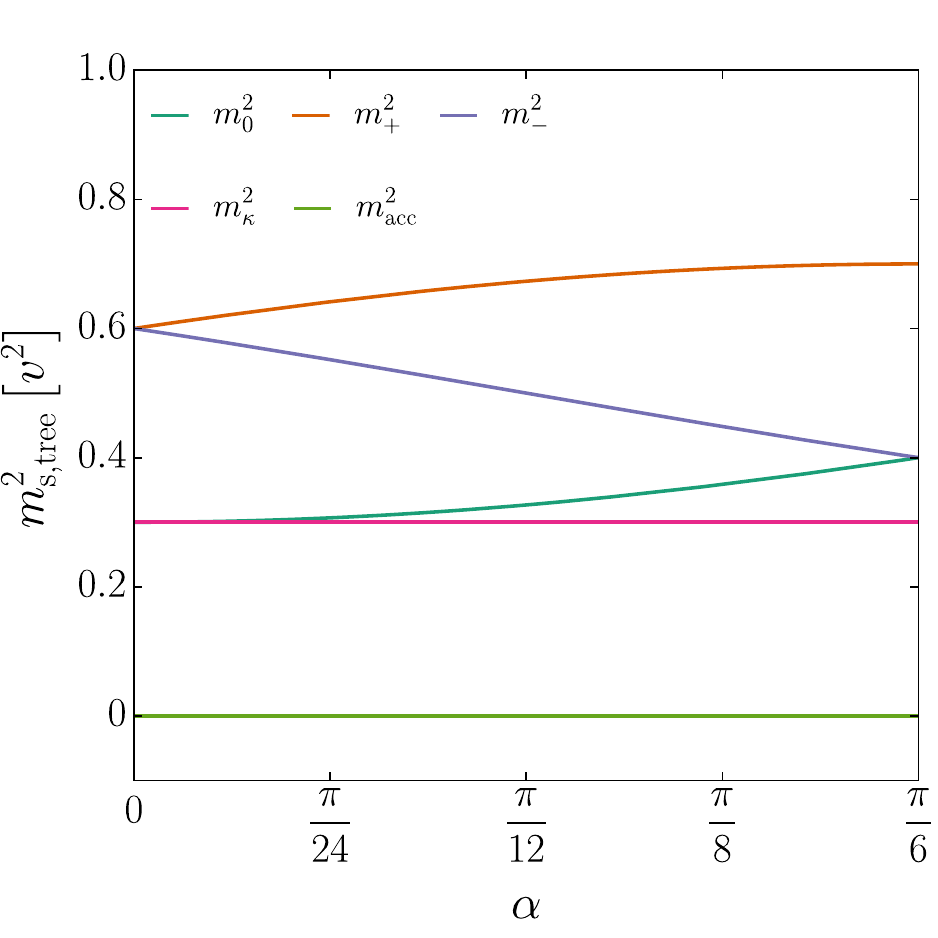}
\qquad\includegraphics[width=.42\textwidth]{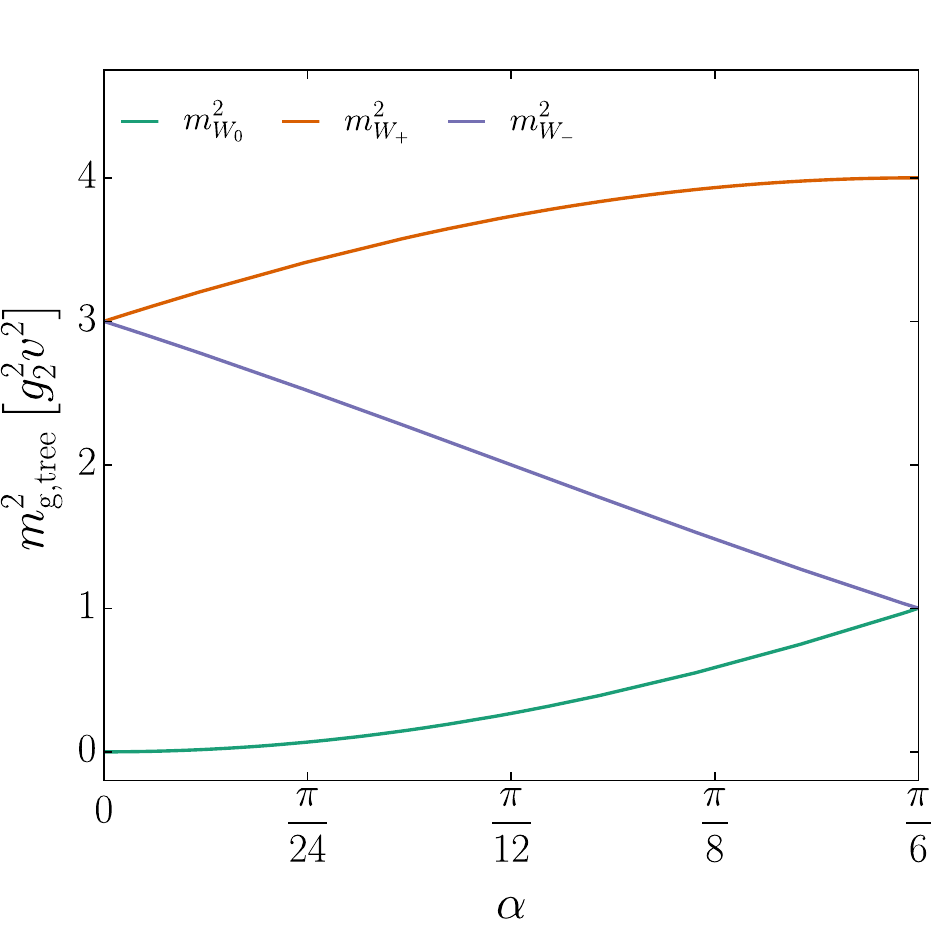}
\caption{Left panel: tree-level masses of the scalar fields moving along the flat direction $\alpha$, for some benchmark values of the couplings, $\kappa = 0.3$ and $\delta = 0.1$. The mass of the radial mode $m_\lambda^2$ is not shown,
as it is controlled by an independent quartic coupling.
Right panel: tree-level masses of the $\SU{2}$ vector bosons in units of $g_2^2 v^2$. At $\alpha = 0$ one of the gauge bosons is massless as $\U{1}'$ is preserved:  at such point the corresponding would-be NGB becomes a second accidentally massless scalar.}
\label{fig:TreeMass}  
\end{center}
\end{figure}

Explicitly, an $\SU{2}\simeq\SO{3}$ five-plet can be represented as a traceless symmetric $3\times 3$ matrix, $\Phi \equiv \phi_j \lambda_j$, with a convenient basis given by the symmetric Gell-Mann matrices $\lambda_{1,3,4,6,8}$.  The $\SO{3}$ transformation $O$ acts as $\Phi\to O\Phi O^T$. Using $\SO{3}\times\U{1}$ invariance, the VEV can thus be chosen real and diagonal:
\be\label{eq:alpha}
\langle\Phi\rangle=\frac{v}{\sqrt{2}}\left(\lambda_3\sin\alpha+\lambda_8\cos\alpha\right)\,,\qquad\lambda_3=\left(\begin{array}{ccc}1&&\\&-1&\\&&0\end{array}\right)\,,\qquad\lambda_8=\frac{1}{\sqrt{3}}\left(\begin{array}{ccc}1&&\\&1&\\&&-2\end{array}\right)\,.
\ee
The accidentally flat direction is parameterised by the angle $\alpha$.
One can show that $\alpha\sim\alpha+\pi/3$ as well as $\alpha\sim-\alpha$ under $\SU{2}\times\U{1}$, 
hence the fundamental domain of $\alpha$ can be chosen to be $\alpha\in[0,\,\frac{\pi}{6}]$. After diagonalisation, the scalar mass matrix becomes
\be\label{eq:scmass}
M^2(\alpha)=\diag\left[m_\lambda^2,\,m_\kappa^2,\,m_0^2(\alpha),\,m_+^2(\alpha),\,m_-^2(\alpha),\,\,0,\,0,\,0,\,0,\,0\right]\,,
\ee
where
\begin{align}
&m_\lambda^2=\lambda\,v^2,&&m_\kappa^2=\kappa\,v^2,\nonumber\\
&m_0^2(\alpha)=\left(\kappa+4\delta \sin^2 \alpha\right)v^2, &&m_{\pm}^2(\alpha)=\left[\kappa+4\delta\sin^2\left(\alpha\pm\frac{\pi}{3}\right)
\right]v^2\,.
\end{align}
with $m_\lambda$ the mass of the radial mode.
When $G$ is gauged, the gauge boson masses are
\be\label{eq:vmass}
m^2_{W0}(\alpha)= 4\,g_2^2\,\sin^2\alpha\,v^2\,,\qquad
m^2_{W\pm}(\alpha) = 4\,g_2^2\,\sin^2\left(\alpha\pm\frac{\pi}{3}\right)v^2\,,\qquad
m_B^2=g_1^2\,v^2\,,
\ee
where $g_1$ and $g_2$ are the $\U{1}$ and $\SU{2}$ gauge coupling, respectively.
Fig.~\ref{fig:TreeMass} shows the correlations among the tree-level masses of scalars
and gauge bosons, as a function of $\alpha$.
Note that the sum of scalar squared masses remain constant along the flat direction, and the same holds for the sum of vector squared masses.

There exists a distinguished point on the vacuum manifold, corresponding to a null eigenvector of one of the $\SU{2}$ generators, $T^3$ say, 
where a $\U{1}'$ subgroup of $\SU{2}$ is unbroken.
In the above parameterisation, it is given by $\alpha=0$.
At this enhanced-symmetry point, there appears a second accidentally massless scalar, which corresponds to the $\U{1}'$ would-be NGB. This should be contrasted with the standard picture in the absence of a flat direction: in that case an enhanced-symmetry vacuum is an isolated minimum of the potential, and an unbroken gauge symmetry implies a massless gauge boson, but no massless scalar. 
There are degeneracies in the massive spectrum too, $m_\kappa=m_0$, $m_+ = m_-$ and
$m_{W+}=m_{W-}$; they are similarly associated with $\U{1}'$-charged states.
The two accidents together form a complex scalar with $\U{1}'$ charge $2$. 

A second distinguished point along the flat direction corresponds to an eigenvector of maximal eigenvalue $2$ of $T^3$, and coincides with the opposite endpoint of the fundamental domain, $\alpha=\pi/6$. 
In this point $G$ is completely broken (as in any generic point), but still there are degeneracies in the mass spectrum, $m_0=m_-$ and $m_{W0}=m_{W-}$.

The extraneous massless degree of freedom which is found at a generic point of the vacuum manifold is not a pseudo-NGB in Weinberg's strict sense \cite{Weinberg:1972fn}, since the potential admits no accidental symmetry it could correspond to. Nevertheless, in the limit where some of the couplings vanish, the symmetry of the potential is enhanced, and some of the scalar degrees of freedom become NGBs.\footnote{In a sense, this is trivially the case for all scalar fields in any model. That is, consider a general model of $N$ real scalars and take the limit where all couplings tend to zero. All scalars then become massless: $N-1$ of them can be regarded as the NGBs of the spontaneous breaking of $\SO{N}$ (the symmetry of the kinetic term) to $\SO{N-1}$ (the subgroup preserved by a generic VEV), while the radial mode associated to the VEV direction can be regarded as the NGB of scale invariance.}

In particular,
    for $\kappa=\delta=0$ and vanishing gauge couplings, the model has a global $\SO{10}$ symmetry spontaneously broken to $\SO{9}$, and indeed there is only one massive radial mode with non-zero mass $m_\lambda$, plus nine NGBs. When $\kappa$ is switched on, the global symmetry is reduced to $\SO{5}\times\U{1}$, cf.~Eqs.~\eqref{V5}--\eqref{bi5}.
This symmetry is spontaneously broken by the VEV to $\SO{4}$, giving rise to five massless NGBs, while four scalars acquire a mass $m_\kappa$. When $\delta$ is switched on, the symmetry is explicitly broken to $\SU{2}\times\U{1}$, but five massless modes remain. Four of them are NGBs of the exact, spontaneously broken symmetries, while the fifth is the accident. The characteristic feature of this model is that the explicit breaking does \emph{not} induce a tree-level mass proportional to $\delta$ for the accident, which remains light (to the extent that the model is perturbative) even when $\delta$ is of the order of the other quartic couplings.

We conclude by discussing an interesting property of the tree-level mass matrices in the presence of accidents. We will argue that in  models with one or more accidentally-flat directions $\alpha_i$, the quantities $\tr M^2$ and $\tr M^4$ are independent of $\alpha_i$ at the tree level, where $M^2=M^2(\alpha_i)$ stands either for the scalar mass matrix, or the vector boson mass matrix $M_W^2$, or the fermion mass matrix $M_F^2\equiv m_F^\dag m_F$. The argument goes as follows: these traces are quadratic or quartic polynomials in the scalar VEV components $v_j$, with all indices contracted in a $G$-invariant way. On the other hand, the scalar potential $V(\phi_j)$ contains, by definition, all independent $G$-invariant polynomials in $\phi_j$ up to degree four. The minimum value of the potential, $V(v_j)$, is constant along the accidentally flat directions $\alpha_i$. This implies that the vacuum is invariant under a symmetry $G_v$ larger than $G$. Let us assume that the action of $G_v$ lifts to a linearly realized action on the initial scalar fields (in the above example, $G_v=\SO{5}\times\U{1})$. Then, the VEVs of $G_v$-invariant operators in the potential are constant as the $\alpha_i$ vary. On the other hand, the $G_v$-breaking operators vanish in the vacuum. In conclusion, there is no $\alpha_i$-dependent polynomial that can contribute to the trace of $M^2(\alpha_i)$ or $M^4(\alpha_i)$. One can check that this argument holds for all the models with accidents considered in this paper.
%Either way, the traces of the mass matrices remain constant along the accidentally-flat directions.}

%%%%%%%%%%%%%%%%%%%%%%%%%%%%%%%%%%%%%%%%%%%%%%%%%%%%%%
\subsection{One-loop lifting of the flat direction}
\label{sec:veff}

Let us focus on the special $\U{1}'$ preserving point of the previous section, at $\alpha=0$. At the one-loop level, the tree-level scalar potential is corrected by the Coleman-Weinberg effective potential \cite{Coleman:1973jx, Jackiw:1974cv}, 
\be
 \Delta V_{\rm CW}=\frac{1}{64\pi^2}\,\text{\rm Str}\,\left({\cal M}^4\log\frac{{\cal M}^2}{\Lambda^2}\right)\,,
\ee
where Str denotes the weighted supertrace, $\Lambda$ is the renormalisation scale, and ${\cal M}^2$ is the scalar-field dependent mass-squared matrix. The one-loop effective potential gives rise to a $\Lambda$-dependent tadpole term for the radial mode of tree-level mass $m_\lambda$. We impose that this tadpole should vanish as a renormalisation condition, i.e.~we define the renormalised VEV to be  $v^2=2\mu^2/\lambda$ in terms of the renormalised $\mu$ and $\lambda$. One then finds that all one-loop tadpole terms for the other scalar fields vanish as well, so that the $\U{1}'$ preserving point remains a critical point of the effective potential. Concerning the mass spectrum, the NGBs remain massless as they should, while the two modes which were accidentally massless at the tree level pick up a finite and positive one-loop mass,
\be
 m_{\text{acc}}^2=\frac{1}{4\pi^2}\left[3\,g_2^2\, m_{W\pm}^2+\delta\,m_{\pm}^2\;f\left(\frac{m_0^2}{m_\pm^2}\right)\right]\Biggr|_{\alpha=0}\,,
\ee
where $m_{W\pm}^2(0)=3\,g_2^2v^2$, $m_\pm^2(0)=(3\delta+\kappa)v^2$,  $m_0^2(0)=\kappa v^2$, and $f$ is a positive-definite function,
\be
f(x)=1-x+x\log x\,.
\ee
We conclude that the symmetry-enhanced point becomes an isolated minimum of the potential, after including one-loop corrections.

For the second distinguished point on the tree-level vacuum manifold, at its opposite end $\alpha=\pi/6$, one finds that there is similarly no tadpole term induced at one loop for the tree-level massless mode (once the one-loop tadpole for the radial mode has been subtracted). However, its one-loop mass is instead tachyonic, and can be written as
\be
 \tilde m_{\text{acc}}^2=-\frac{1}{4\pi^2}\left[3\,g_2^2\,m_{W-}^2\;f\left(\frac{m_{W+}^2}{m_{W-}^2}\right)+\delta\,m_-^2\;f\left(\frac{m_+^2}{m_-^2}\right)\right]\Biggr|_{\alpha=\frac{\pi}{6}}\,,
\ee
with $m_{W+}^2(\frac{\pi}{6})=4\,g_2^2 v^2$, $m_{W-}^2(\frac{\pi}{6})=g_2^2 v^2$, $m_+^2(\frac{\pi}{6})=(\kappa+4\delta)v^2$, and $m_-^2(\frac{\pi}{6})=(\kappa+\delta)v^2$. Therefore this second point is a saddle point of the effective potential. 

Note that the accident effective potential does not depend on $g_1$ at one loop. Indeed, gauging $\U{1}$ preserves the $\SO{5}$ symmetry which is recovered for $\delta\into 0$ and $g_2\into 0$. In this limit the accidents become NGBs of $\SO{5}/\SO{4}$, and therefore their effective potential must vanish.

%%%%%%%%%%%%%%%%%%%%%%%%%%%%%%%%%%%%%%%%%%%%%%%%%%%%%%%%%%%%%%
\subsection{Coupling to fermions: accident misalignment 
}\label{subsec:abelianhiggs}

The low-energy fluctuations around the $\U{1}'$-preserving point at $\alpha=0$ are described by a simple effective field theory: a $\U{1}'$ gauge theory with one light charged scalar. The scalar mass $m_{\text{acc}}^2$ is loop-suppressed with respect to the masses of the heavier states constituting the UV completion. It is interesting to study the question whether the symmetry-enhanced point can be destabilized by loop effects, which would spontaneously break the residual $\U{1}'$. However, we have shown that, with a field content of only scalars and gauge bosons, $m_{\text{acc}}^2$ is always positive. 

\begin{figure}[ht]
 \centering
 \includegraphics[width=.65
\textwidth]{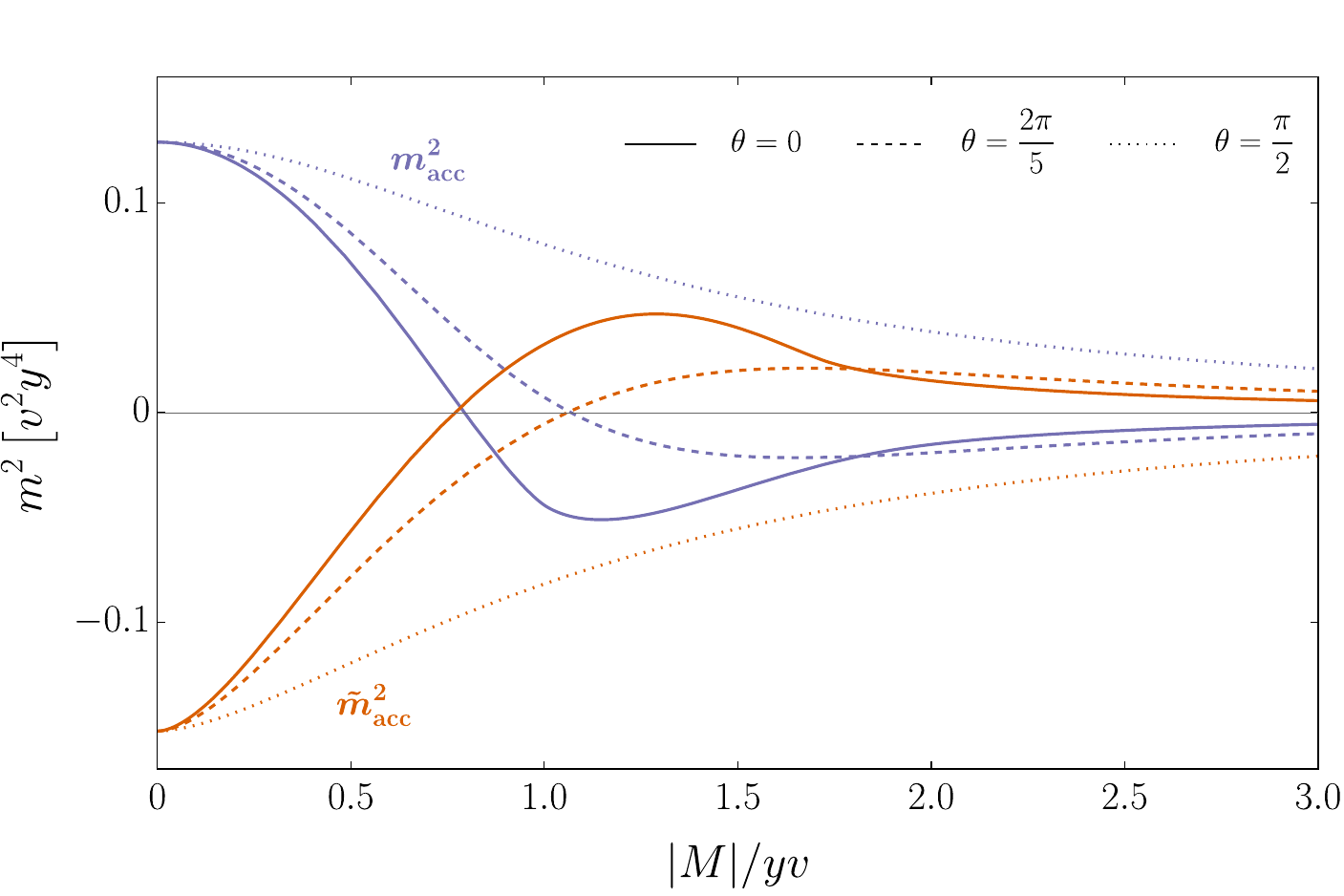}
 \caption{The fermion contribution to the one-loop mass-squared of the accident,  at $\alpha=0$ (purple curves) and $\alpha=\pi/6$ (orange curves), as a function of $|M|/yv$, where $M = \lvert M \rvert e^{i\theta}$ is the Dirac mass and the two Yukawa couplings are assumed to be equal, $y_\chi=y_\psi=y$. For e.g.~$\theta=0$ (solid curves), the $\U{1}'$-preserving point is destabilised for $M\simeq 0.8\, yv$, whereupon the $\U{1}'$-breaking point at $\alpha=\pi/6$ becomes the minimum.}\label{fig:yukmass}
\end{figure}

We therefore add to the model a minimal anomaly-free set of fermion fields coupled to the five-plet $\phi$. We take $\chi$ and $\psi$ to be left-handed Weyl fermions, transforming as $\bf{3_{\pm 1/2}}$ with respect to $\SU{2}\times \U{1}$,\footnote{
Another minimal choice would be $\chi\sim \bf{5_0}$, $\psi\sim \bf{1_{-1}}$, $\bar\psi\sim \bf{1_{1}}$. However in this case the Yukawa couplings are $SO(5)$ symmetric, therefore they induce no potential for the accidents.} which allows for the terms
\be\label{eq:Lferm}
{\cal L}\supset y_\psi\,\psi^T\Phi \psi
+y_\chi\,\chi^T \Phi^*\chi
+M\, \psi^T\chi\text{+ h.c.}\,,
\ee
where the matrix $\Phi$ was defined above \eq{eq:alpha}. The complex phases of the Yukawa couplings $y_\psi$ and $y_\chi$ can be set to zero without loss of generality, but then the phase of the Dirac mass $M$ becomes physical. In Fig.~\ref{fig:yukmass} we plot the accident mass at $\alpha=0$ induced by fermion loops, in the limit where gauge and scalar self-couplings are negligible, for the special case $y_\chi=y_\psi\equiv y$. For a small $|M|\ll yv$, we find $m_{\rm acc}^2$ is always positive, while for sizeable $|M|$ with a suitable phase, there is a region where the accident becomes tachyonic, so that fermion loops do indeed destabilize the symmetry-enhanced point. In this region, the former saddle point $\alpha=\pi/6$ becomes the minimum of the effective potential, and $\U{1}'$ is spontaneously broken at the loop level.

Note that, even though this spontaneous breaking is induced at one-loop, its scale is the same as for the tree-level breaking of $\SU{2}$, that is to say, the gauge boson masses in the new minimum are all of the order $g_2\,v$. In other words, the effective field theory valid close to the symmetry-enhanced point is not suitable anymore, because the new $\U{1}'$-breaking minimum is at distance of the order of the cutoff $v$ in field space. It would be more interesting to find a configuration where the effective potential has a minimum $\alpha_{\rm min}$ close to (but not exactly at) the $\U{1}'$-preserving point, which would imply a breaking scale $v'\ll v$, where we define $v'/v\equiv\sin(3\alpha_{\rm min})$. In the effective theory well below the scale $v$, one could then identify the accident field with the Higgs boson of $\U{1}'$ breaking. 

To this end, we now discuss the combined effect of scalar self-couplings, gauge couplings and Yukawa couplings. The one-loop effective potential along the accident direction can be evaluated explicitly as a function of $\alpha$ because, on the tree-level vacuum manifold, ${\cal M}^2$ is simply given by the tree-level mass matrix which is easily diagonalised.
Subtracting the radial tadpole and the vacuum energy at the renormalisation point $\alpha=0$, one obtains
\be\begin{split}\label{eq:VCW}
\Delta V_{\rm CW}(\alpha)\Bigr|_{\text{tree-level vacuum}}=\frac{1}{64\pi^2}\Biggl[&\tr\left( M^4(\alpha)\log \frac{M^2(\alpha)}{\Lambda^2}-M^4(0)\log \frac{M^2(0)}{\Lambda^2}\right)\\
+&\,3\,\tr \left( M_W^4(\alpha)\log \frac{M_W^2(\alpha)}{\Lambda^2}- M_W^4(0)\log \frac{M_W^2(0)}{\Lambda^2}\right)\\
-&\,2\,\tr\left( M_F^4
(\alpha)\log \frac{M_F^2(\alpha)}{\Lambda^2}- M_F^4(0)\log \frac{M_F^2(0)}{\Lambda^2}\right)\Biggr]\,.
\end{split}
\ee
Here the scalar mass matrix is given by Eq.~\eqref{eq:scmass}, and the eigenvalues of the gauge boson mass matrix $M_W^2$ by Eq.~\eqref{eq:vmass}. The $6\times 6$ fermion mass matrix $m_F$ resulting from Eq.~\eqref{eq:Lferm} can likewise be diagonalised analytically, and we denoted $M_F^2\equiv m_F^\dag m_F$. Note that the $\Lambda$-dependence in Eq.~\eqref{eq:VCW} is spurious, since $\tr M^4(\alpha)$, $\tr M_W^4(\alpha)$ and $\tr M_F^4(\alpha)$ are $\alpha$-independent (see Section \ref{potT}): the divergences have already been subtracted, and the resulting effective potential is finite.

\begin{figure}[ht]
\begin{center}
\includegraphics[width=.42\textwidth]{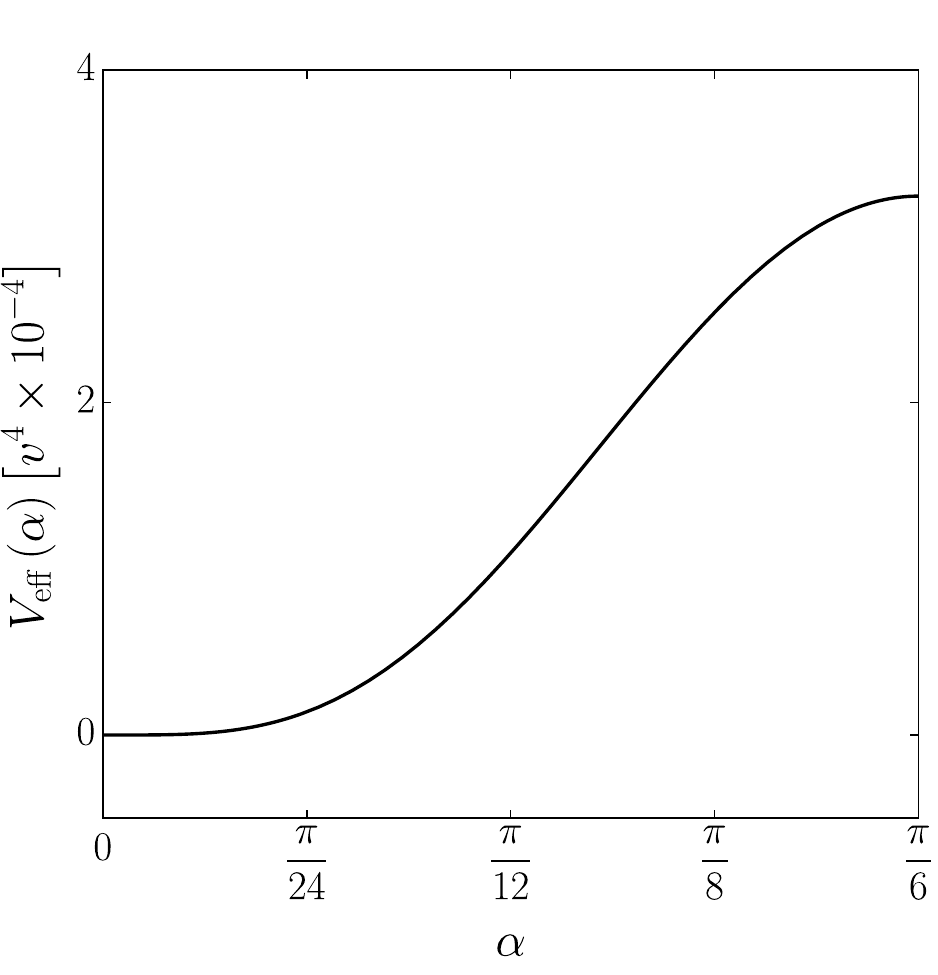}
\qquad\includegraphics[width=.42\textwidth]{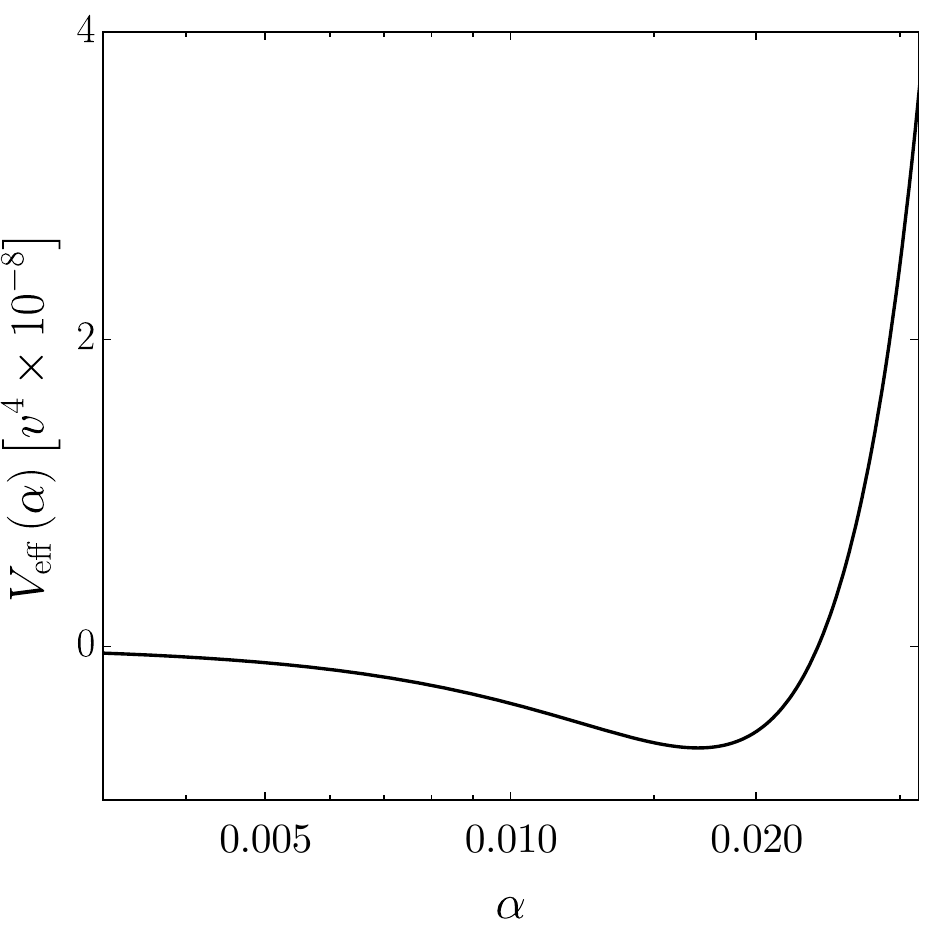}
\caption{Left panel: one-loop effective potential along the accident direction $\alpha$, for $\kappa = 1$, $\delta = 0.995$, $\theta = 0$, $\left| M \right|/v= 0.8$, $y_{\chi}=y_{\psi}=1$ and negligible gauge couplings. Right panel: a zoom of the left-hand panel in the region close to the origin, showing the minimum at $\alpha
\simeq 0.02$.} 
\label{fig:Veff}  
\end{center}
\end{figure}

The couplings can now be chosen to obtain a minimum of the one-loop corrected potential parametrically close to $\alpha=0$. The potential for an example point in parameter space is shown in Fig.~\ref{fig:Veff}. At this point, the gauge couplings are negligible, and the loop contributions from scalars and fermions largely balance each other, leading to a small misalignment from the $\U{1}'$-preserving direction, with $v'/v \simeq 0.06$.
However, such a minimum can be obtained only at the price of fine-tuning the parameters at the per-mille level. This in turn implies that higher loop corrections could significantly shift the minimum of the potential. To understand the fine-tuning, it is instructive to consider the Fourier expansion of the contributions to the effective potential. One has
\be
\Delta V_{\rm CW}(\alpha)\Bigr|_{\text{tree-level vacuum}} \simeq c_6 \cos \left(6 \alpha \right) + c_{12} \cos \left(12 \alpha \right)\,,
\ee
where $c_6$ and $c_{12}$ are functions of the couplings. Both the fermionic and  bosonic contributions to $c_{12}$ turn out to be numerically suppressed with respect to the respective contributions to $c_6$, by at least two orders of magnitude (and higher harmonics are even more suppressed, which is why they are neglected here).\footnote{To be precise, this statement holds in the fermion sector for all values of the couplings that tend to destabilise the $\alpha=0$ vacuum, see Fig.~\ref{fig:yukmass}.} For generic values of the couplings, the effective potential is therefore dominated by the lowest harmonic, $\Delta V_{\rm CW}\simeq c_6 \cos(6\alpha)$, and the minimum is either at $\alpha=0$ or at $\alpha=\pi/6$ depending on the sign of $c_6$. In order to obtain a small misalignment, the fermionic and bosonic contributions to $c_6$ must cancel for the most part, and moreover $c_6/c_{12}$ must be accurately tuned, in order to achieve $(v'/v)^2 \simeq  1/2 + c_6/(8 c_{12})\ll 1$. The tuning in the coefficients is numerically of the order of higher-loop effects, which we neglected in our analysis. In other words, the exact position of the minimum shown in Fig.~\ref{fig:Veff} is not under theoretical control.

%%%%%%%%%%%%%%%%%%%%%%%%%%%%%%%%%%%%%%%%%%%%%%%%%%
\subsection{A supersymmetric model with accidents}

The insights gained from the $\SU{2}\times\U{1}$ model of section \ref{potT} can be used to build a simple supersymmetric model of accidentally light fields. Promote $\phi$ to a chiral supermultiplet $\Phi=\varphi+\theta\psi+\theta^2 F$ of $N=1$ SUSY, and include a second chiral supermultiplet $\widetilde\Phi=\tilde\varphi+\theta\tilde\psi+\theta^2\tilde F$ with the conjugate quantum numbers ${\bf 5}_{-1}$. Introduce the superpotential
\be
W=-\mu\,S+\frac{1}{2M}\left[\lambda\,SS+\kappa\,(SS-S' \widetilde{S}')+\delta A^a A^a\right]
\ee
with the bilinear chiral superfields $S=\Phi\widetilde\Phi$, $S'=\Phi\Phi$, 
$\widetilde S'=\widetilde\Phi\widetilde\Phi$, and $A^a=\widetilde\Phi T^a\Phi$. This is the most general superpotential compatible with $G=\SU{2}\times\U{1}$, up to mass dimension four. If $G$ is a global symmetry, the full scalar potential is given by the $F$-term potential,
\be
V_{F}=\left.K^{I\bar J}W_I W^*_{\bar J}\right|_{\Phi\into\varphi,\widetilde\Phi\into\tilde\varphi}\,,
\ee
where subscripts indicate derivatives with respect to superfields, and $K^{I\bar J}$ is the inverse K\"ahler metric. 

Any critical point with $W_I=0$ for all $I$ gives a SUSY vacuum.
By comparing $W$ with the non-supersymmetric scalar potential $V$ of \eq{V5}, one observes that SUSY vacua are in one-to-one correspondence with the critical points of $V$, and are obtained from the latter by simply replacing $\mu^2\into \mu M$, $\phi\into\varphi$ and $\phi^*\into\tilde\varphi$. This is despite the fact that the supersymmetric scalar potential $V_F$ is not at all similar to the non-supersymmetric $V$; in particular, it depends on twice as many scalar fields and includes operators up to dimension 6. Thus, following the analysis of section \ref{potT}, in the region where all couplings are real and positive there are SUSY vacua corresponding to \eq{minV}: the associated moduli space is compact, with no runaway directions; the VEVs of $\varphi$ and $\tilde\varphi$ are aligned; $G$ is completely broken at a generic point in moduli space, and broken to $\U{1}'$ at a special point; there are five massless superfields, four of which are Goldstones while one is an accident, associated to the compact modulus parameterised as in \eq{eq:alpha}; at a special point in moduli space there is one less Goldstone superfield and one more accident superfield.

When $G$ is gauged, the $D$-terms,
\be
D^a=g_2(\varphi_i^* T^a_{ij}\varphi_j+\tilde\varphi_i^* T^a_{ij}\tilde\varphi_j)\,,\qquad 
D^0=g_1(|\varphi|^2-|\tilde\varphi|^2)\,,
\ee
must also vanish in a SUSY vacuum. This is the case if the VEVs of the five-plets $\varphi$, $\tilde\varphi$, $\varphi^*$ and $\tilde\varphi^*$ are all aligned in field space, and $|\vev{\varphi}|=|\vev{\tilde\varphi}|$. The erstwhile Goldstone chiral superfields are absorbed by the gauge superfields. One chiral supermultiplet remains massless on the entire moduli space, and a second one becomes massless at the special point. Thus, the number of chiral-superfield accidents matches the number of 
real-scalar accidents of the non-supersymmetric model.

The interest of this model is that a SUSY accident, i.e.~a tree-level massless supermultiplet, remains massless at all orders in perturbation theory by the non-renormalisation theorem, as long as SUSY is exact. In section \ref{susyapp} we will sketch possible phenomenological applications of SUSY accidents.

%%%%%%%%%%%%%%%%%%%%%%%%%%%%%%%%%%%%%%%%%%%%%%%%
\section{The first in a series of accidents: an SU(3) ten-plet}\label{sec:su3}
%%%%%%%%%%%%%%%%%%%%%%%%%%%%%%%%%%%%%%%%%%%%%%%%

Accidentally massless scalars are not specific to the $\SU{2}\times\U{1}$ five-plet model. The same phenomenon occurs in other models with large representations. Consider for instance the symmetry group $G=\SU{N}\times\U{1}$ with a scalar field $\phi^{ijk}$ in the three-index symmetric representation of $\SU{N}$ and unit $\U{1}$ charge. Here $i,j,k=1,\dots,N$ are fundamental indices; the $\phi$ multiplet contains $N(N+1)(N+2)/6$ components. For $N=2$, one can check that the most general renormalisable scalar potential compatible with $G$ gives tree-level masses to all scalars, while accidents start appearing for $N\geq 3$. In the following we give some details on the $N=3$ model, commenting on $N>3$ at the end of the section.

Let us thus consider $G=\SU{3}\times\U{1}$ with a scalar field $\phi\sim{\bf 10}_1$. Its components can be alternatively written as a vector with one index in the ten-dimensional representation, $\phi^I$ with $I=1,\dots,10$. The scalar potential for a ${\bf 10}$ representation of $\SU{3}$ has been partially analysed previously, e.g.~to derive discrete flavour symmetries \cite{Luhn:2011ip} or to stabilise a scalar dark matter candidate \cite{Frigerio:2022kyu}. In the latter paper, the existence of tree-level flat directions was pointed out. Here we investigate in more detail their nature and implications. 

The most general renormalisable potential invariant under  $\SU{3}\times\U{1}$  includes, besides the mass term, only two algebraically independent quartic invariants, 
\be\label{eq:V10}
V=-\mu^2\,S+\frac12\left(\lambda\, S^2+\delta\, A^a A^a\right)\,,
\ee
where the singlet and adjoint bilinears are defined by
\be
S=\phi^\dag\phi\,,\qquad\qquad A^a=\phi^\dag T^a\phi\qquad(a=1,\ldots 8)\,.
\ee
Here $T^a$ are the $\SU{3}$ generators in the $\bf{10}$ representation, satisfying $\tr(T^a T^b) =15/2\, \delta^{ab}$. The $\SO{20}$ global symmetry of a free theory of 20 mass-degenerate real scalars is respected by the $\lambda$ term, however it can be checked that $\delta$ breaks it explicitly to $\SU{3}\times\U{1}$, with no larger continuous symmetry surviving. We take $\mu^2>0$ to realise spontaneous symmetry breaking, while 
boundedness from below requires $\lambda>0$ and $\delta > -\lambda/3$. It so happens that accidentally massless scalars arise only for positive values of $\delta$, hence we focus on $\delta>0$.

Since only the $A^a A^a$ term is sensitive to the direction of the VEV in field space, the potential is minimized in a direction where $A^a A^a$ is minimal, i.e.~where $\vev{A^a} =0$. The VEV can then be rescaled to satisfy $\vev S=\mu^2/\lambda\equiv v^2/2$, in order to obtain a minimum of the full potential. One such direction can readily be identified as the common null eigenvector of the two Cartan generators of $\SU{3}$, conventionally taken to be $T_3$ and $T_8$, so that $\SU{3}$ breaks to $\U{1}_3\times \U{1}_8$. We call the corresponding direction in field space  point (I), with $\vev\phi=v_{\rm (I)}/\sqrt{2}$. In terms of the  $\phi^{ijk}$ components, $G$ transformations can be used to bring it to the form $\vev{\phi^{(123)}}=|v|/\sqrt{2}$, with all other components VEVs vanishing. Here the parentheses in $\phi^{(123)}$ stand for weighted symmetrisation. 

At point (I), out of the 20 real scalar fields, only seven are massive and 13 are massless at the tree level, despite the fact that only seven generators are spontaneously broken. The remaining six massless scalar fields are accidents. The diagonal form of the scalar mass matrix is
\be
M^2_{\rm(I)}=\diag\left(m_\lambda^2,\,6\times m_\delta^2\,,13\times 0\right)\,,
\ee
where the masses of the radial mode and the other massive modes are given by $m_\lambda^2=\lambda\,v^2$ and $m_\delta^2=\delta\,v^2$, respectively. When $\SU{3}\times\U{1}$ is gauged with couplings $g_3$ and $g_1$, the gauge boson masses are $m_V^2= g_3^2\,v^2~(\times\, 6)$, $m_{V_3}^2=m_{V_8}^2=0$ and $m_B^2=g_1^2\,v^2$. It can be checked that there are no points preserving a larger continuous symmetry than $\U{1}_3\times \U{1}_8$ on the vacuum manifold, and that the symmetry-preserving point is unique up to $G$-transformations.

Away from the special point (I), at a generic minimum defined by $\vev{A^a}=0$ and $\vev S=v^2/2$, we find that $G$ is completely broken, with nine NGBs, nine massive modes, and two accidentally massless modes. To explicitly construct the associated flat directions, consider a second point (II) in field space, $\vev\phi=v_{\rm (II)}$, defined by $\vev{\phi^{111}}=\vev{\phi^{222}}=\vev{\phi^{333}}=|v|/\sqrt{6}$ and all other VEVs vanishing. The points (I) and (II) are not equivalent under a $G$-transformation, as evident from  the different tree-level mass spectrum:
\be
M^2_{\rm (II)}=\diag\left(m_\lambda^2,\,6\times \frac{m_\delta^2}{2},\,2\times \frac{3m_\delta^2}{2},\,11\times 0\right)~.
\ee
Likewise, the $\SU{3}$ gauge boson masses at point (II) are $m^2_{V}/2~(\times 6)$ and $3 m_{V}^2/2~(\times 2)$.

In fact, any superposition of the VEV directions defining points (I) and (II) gives rise to a minimum of the potential with $\vev{\phi}=v_{\rm(I)}\,\cos\alpha+v_{\rm(II)}\sin\alpha$, where $\alpha\in[0,\pi/2]$.
Moreover, it can be checked that the vacuum energy does not depend on the complex phases of the four VEVs of $\phi^{(123)}$ and $\phi^{iii}$. Three of them can be rotated away using the $\U{1}_3\times \U{1}_8\times \U{1}$ generators. However, the fourth phase $\beta$ is physical, and so one obtains a two-parameter family of physically inequivalent vacua:
\be
\vev{\phi^{111}}=\vev{\phi^{222}}=\vev{\phi^{333}}=\frac{|v|}{\sqrt{6}}\sin\alpha\,,\qquad\vev{\phi^{(123)}}=\frac{|v|}{\sqrt{2}}\,\cos\alpha\;e^{i\beta}\,,\qquad\text{all other VEVs }=0\,.
\label{vacuum}\ee

To summarize, the manifold of tree-level degenerate vacua is eleven-dimensional. At generic points in this manifold, $G$ is completely broken, and nine of the flat directions are Goldstone directions. When $G$ is gauged, gauge-equivalent points are identified, and nine NGBs are absorbed by the gauge bosons. There remains a two-dimensional manifold of gauge-inequivalent vacua, corresponding to two accidentally massless scalars, parameterised by two angles $\alpha$ and $\beta$. At point (I) the vacuum manifold degenerates, there is a residual symmetry $\U{1}_3\times \U{1}_8$, and four additional tree-level accidents appear. Two of them correspond to the would-be NGBs of restored symmetries, while the other two are of different nature; they are not associated to any of the 11 flat directions, but arise only at the special point.

The non-vanishing scalar masses along the vacuum manifold can be written as the roots of a certain cubic polynomial; their explicit expressions are not very illuminating. Nevertheless, they obey a simple sum rule, as is easily checked directly from \eq{eq:V10}: $\tr M^2=\left(\lambda+C_2\,\delta\right)v^2$, where $C_2=6$ is the quadratic Casimir. Similarly, the sum of gauge boson masses squared is constant and equal to $(C_2\,g_3^2+g_1^2)v^2$; see Section \ref{potT}.

Let us focus on the special point (I), where $\alpha=0$ and where $\beta$ becomes the Goldstone direction of the spontaneously broken $\U{1}$. The six non-Goldstone flat directions correspond to the real and imaginary parts of $\phi^{111}$, $\phi^{222}$ and $\phi^{333}$. The fate of these accidents beyond the tree level can be determined by computing the one-loop effective potential. In analogy with our computation of section \ref{sec:veff}, one finds that subtracting the tadpole term induced for the radial part of $\phi^{(123)}$ renders the one-loop masses finite. All six accidents receive a positive and equal one-loop mass given by
\be
m_{\text{acc}}^2=\frac{3}{16 \pi^2}\left( \delta\,m_{\delta}^2+3g_3^2\, m_V^2\right)\,.
\ee
The flat directions are therefore lifted, and the symmetry-enhanced point (I) becomes an isolated minimum of the effective potential, i.e.~the physical vacuum of this model. Point (II), on the other hand, is found to be destabilised by loops and becomes a saddle point of the effective potential.

The mass degeneracy of the six accidents (as well as the other mass degeneracies in the spectrum) are due to discrete symmetries preserved after spontaneous symmetry breaking. In particular, $G$ contains an $S_3$ subgroup acting by permutation on the $\phi^{ijk}$ indices, left unbroken in the vacuum of \eq{vacuum}. Note the remnant symmetries $\U{1}_3\times \U{1}_8$ and $S_3$ do not commute. For example, the accident components $\phi^{iii}$ carry different charges, yet they form a triplet under permutations. 
In principle, such remnant gauge symmetries could be radiatively broken by fermion loops, which might destabilise the special point, in the spirit of section \ref{subsec:abelianhiggs}; in this case finding the new minimum requires a multi-field effective potential computation.

Finally, let us comment on the analogous models obtained by replacing $\SU{3}$ with $\SU{N}$. The structure of the most general renormalisable potential $V$ is the same as in \eq{eq:V10}, for any $N$. For $\mu^2,\lambda,\delta>0$, the dimension of the tree-level vacuum manifold rapidly grows, yielding $N(N-1)(N-2)/3$ non-Goldstone flat directions.
As for $N=3$, there can be special points on the tree-level vacuum manifold where a subgroup of $G$ remains unbroken, and the number of accidents is enhanced. A systematic classification of these models and their vacua is left for future work.

%%%%%%%%%%%%%%%%%%%%%%%%%%%%%%%%%%%%%%%
\section{Possible applications}\label{sec:apps}

Could accidentally light scalars play a role in real-world particle physics or cosmology? Let us present a few possible applications of phenomenological interest.

%%%%%%%%%%%%%%%%%%%%%%%%%%%%%%%%%%%
\subsection{Accident dark matter}

The toy models presented above could actually describe a realistic dark sector with accidents playing the role of dark matter (DM) candidates. Assume the new scalar field $\phi$ to be a Standard Model (SM) singlet, and the SM sector to be neutral with respect to the dark-sector symmetry $G$. The two sectors can communicate (and thermalise) through a Higgs portal interaction, $\lambda_{H\phi}(H^\dag H)(\phi^\dag \phi)$ with $H$ the SM Higgs doublet. Let us consider the minimal model, with dark gauge symmetry $\SU{2}\times\U{1}$ spontaneously broken to $\U{1}'$. Indeed, we have shown that such symmetry-enhanced vacuum is selected, once the accident flat direction is lifted by radiative corrections (at least in the absence of dark fermions).

This remnant symmetry guarantees that the lightest charged particle is stable (without the need to assume an ad-hoc global symmetry). This is naturally the complex scalar accident, since it is charged under the unbroken $\U{1}'$ and its mass is generated only at loop level.\footnote{Let us remark that, even when fermion loops select a different vacuum, with no continuous unbroken symmetry, there are typically remnant discrete symmetries which also guarantee the accident stability. For example, when the minimum occurs at the second special point with $\alpha=\pi/6$, one can show that the accident is odd under a residual $Z_2$ symmetry, which also explains the mass degeneracies observed in Fig.~\ref{fig:TreeMass}. Therefore, accidents can be good DM candidates even in the absence of $\U{1}$ factors.}

The unbroken gauge symmetry at the special point guarantees the presence of massless dark photons, into which the accidents can annihilate. The DM phenomenology in the case of annihilation into dark, massless gauge bosons is discussed e.g.~in section 5.2.1 of \cite{Frigerio:2022kyu}. In our case, the observed relic density can be reproduced when these annihilations freeze out, as long as $Q_D^4 \alpha^2_{D} \simeq 2.2 \cdot 10^{-10} (m_{ DM}/\hbox{GeV})^2$, where $\alpha_{D} = g_D^2/(4\pi)$ with $g_{D}$ the 
$\U{1}'$ dark gauge coupling and $Q_D=2$ the accident charge. Alternatively, DM annihilation through the Higgs portal can dominate. In this case direct detection constraints require $m_{DM}\gtrsim 2$-$3$~TeV or $m_{ DM}$ within a very narrow window around the resonance $m_{DM}\simeq m_h/2$, see e.g.~\cite{Frigerio:2022kyu}. The DM direct detection, indirect detection and collider constraints all depend on the size of the Higgs portal coupling, which can thus be tested.

A massless dark photon contributes to the  extra-radiation parameter $\Delta N_{\rm eff}$, which is constrained both by BBN and CMB. Therefore, it needs to decouple from the early-universe thermal bath sufficiently early (so that its contribution is sufficiently diluted by the SM reheating, from the decoupling of the various SM particles). 
For a single dark photon, one finds that the decoupling temperature should be above a few hundreds of MeV, which implies a DM mass above a few GeV \cite{Frigerio:2022kyu}. Future CMB observatories are expected to have the ability to rule out or establish an extra radiation component at the level of a single dark photon. Another relevant constraint comes from galactic-scale structure formation: the so-called ellipticity constraint gives an upper bound on the strength of the dark-photon long-range force, $Q_D^2\alpha_D\lesssim 0.4 \sqrt{10^{-11}(m_{DM}/\hbox{GeV})^3}$ \cite{Feng:2009mn,Agrawal:2016quu, McDaniel:2021kpq}, which combined with the relic density constraint implies a DM mass above $\sim 100$~GeV in our case. 

If the Higgs portal coupling is tiny, the SM and dark sectors do not thermalise with each other in the early Universe, but still each sector can thermalise individually. Such a scenario with  two thermal baths leads to a different, but perfectly viable, DM phenomenology \cite{Chu:2011be,Hambye:2019dwd}. In particular if the dark sector has a temperature $T'$ smaller than the one of the visible sector $T$, the dark photon contribution to $\Delta N_{\rm eff}$ is suppressed by a factor of $(T'/T)^3$, and becomes irrelevant.

An analogous analysis holds for the $\SU{3}\times \U{1}$ model: the natural DM candidate is the multiplet formed by the six degenerate accidents, charged under the remnant gauge symmetry $\U{1}_3\times \U{1}_8$, corresponding to two dark photons
(see \cite{Frigerio:2022kyu} for quantitative constraints).

%%%%%%%%%%%%%%%%%%%%%%%%%%%%%%%%%%%%%%%%%%
\subsection{Cosmology along the accident potential}

As well known, any inflationary potential must be extremely flat along the inflaton scalar field direction. The possibility to invoke a shift symmetry to protect the flatness of the inflationary potential has been considered extensively since decades. This constitutes the ``natural inflation'' scenario \cite{Adams:1992bn} (also appearing in axion setups). In these scenarios the inflaton is the NGB of a spontaneously broken continuous symmetry. Inflation requires that the shift symmetry is slightly broken for the potential not to be totally flat, and the inflaton is therefore a pseudo-NGB.

It is interesting to note that the $\SU{2}\times \U{1}$ potential we obtain along the accident direction is of similar form, i.e.~$V = \Lambda^4[1+ a\, \cos (\varphi/v)]$, with $\varphi$ the accident field. This potential is known to be in tension with Planck data \cite{Planck:2018jri}. To be in agreement with Planck data one possibility is to let the inflaton slow roll down this potential and to end inflation by an extra waterfall scalar field, see e.g.~\cite{Ross:2016hyb}. However, such a field does not exist in our model, and adding it would imply to significantly change the structure of the scalar potential and of its minima. A better option might be to invoke a modified cosmology posterior to inflation (see e.g.~\cite{Stein:2021uge}), or non-minimal couplings to gravity (see e.g.~\cite{Reyimuaji:2020goi}). 

In general, the potential of accidents can have a richer structure than the minimal expression above: firstly, it may include higher harmonics of the form $a_n\cos(n\,\varphi/v)$; secondly, in models with multiple tree-level flat directions as the $\SU{3} \times \U{1}$ model, slow roll may occur in the corresponding multi-field space. A dedicated study would be needed to assess the implications of these features on inflation.

Note also that, as the accident oscillates around the bottom of its potential, it triggers a burst of (massless dark photon) particle production whenever it crosses the enhanced-symmetry point \cite{Kofman:1997yn, Kofman:1994rk}. This possibility of producing dark vector bosons could have interesting consequences for cosmology, see e.g.~\cite{Dror:2018pdh} in a somewhat different context.

We conclude by commenting on the possibility to have a first-order cosmological phase transition along the accident direction.\footnote{We thank an anonymous Referee for inquiring about this possibility.} 
We showed in section \ref{subsec:abelianhiggs} that the accident effective potential can develop a minimum away from the $\U{1}'$-preserving point at $\alpha=0$, so that $\U{1}'$ is spontaneously broken.
At sufficiently large temperatures, thermal corrections will dominate the effective potential, and tend to restore the symmetry with a minimum at $\alpha=0$. As the Universe expands, the temperature $T$ decreases, and the potential develops a second minimum which may be separated by a barrier from the one at the origin. This is typically the case if the tree-level potential at zero temperature is flat, and the flat directions are lifted radiatively by the one-loop effective potential. 
%Then the phase transition is of first order, since it can proceed only via tunneling or thermal fluctuations. 
Then the phase transition can be of first order, proceeding via tunneling (this ends a period of supercooling, during which the scalar field lies in the false minimum and dominates the Universe energy density, see e.g.~\cite{Iso:2017uuu,vonHarling:2017yew,Hambye:2018qjv}).
As $T$ further decreases, the origin becomes a local maximum and, at $T=0$, we recover the potential shown in Fig.~\ref{fig:Veff}.
First-order phase transitions may be relevant for baryogenesis \cite{Kuzmin:1985mm,Morrissey:2012db}, production of primordial black holes (see \cite{Green:2020jor} and references therein), and/or for a signal of stochastic gravitational wave background (see \cite{Maggiore:2018sht} for an overview).

%%%%%%%%%%%%%%%%%%%%%%%%%%%%%%%%%%%%%%%%%%%%%%%
\subsection{The Higgs as an accident}

The only elementary scalar field in the SM is the Higgs boson, which at $125$ GeV is much lighter than the scale of new physics. The absence of new physics up to the (multi-)TeV scale constitutes the ``little hierarchy problem''. Ultraviolet embeddings of the SM predicting a loop-suppressed Higgs mass are therefore appealing, and have been widely explored. The accident mechanism may lead to an additional class of such models.

To address the little hierarchy problem with accidents, one would need to identify a model where, at some  special point in the tree-level vacuum manifold, the remnant symmetry contains the electroweak symmetry $\SU{2}_w\times \U{1}_Y$ (or even the entire custodial symmetry of the SM Higgs potential), and the accidentally light scalars transform as an electroweak doublet. We have shown in section \ref{subsec:abelianhiggs} how radiative corrections may give a VEV to the accidents, thus breaking the remnant symmetry at a scale parametrically smaller than the initial scale  of spontaneous symmetry breaking. The accident models which we have identified so far feature only abelian $\U{1}^n$ remnant symmetries, but this is likely a consequence of the simplest possible choices for the symmetry $G$ and for the representation of $\phi$. 

The phenomenology of an accidental Higgs would resemble that of composite pseudo-NGB Higgs models (see \cite{Bellazzini:2014yua,Panico:2015jxa} for reviews) or little-Higgs models (see \cite{Schmaltz:2005ky} for a review). In the little-Higgs scenario, the Higgs mass is loop-suppressed because it is protected by a large global symmetry, explicitly broken only by the product of at least two independent couplings, typically gauge or Yukawa couplings. In the accident scenario, an enlarged symmetry is broken by a single scalar quartic coupling, nonetheless the accident mass is loop-suppressed thanks to the restrictive structure of the scalar potential. In composite Higgs models, it is typically assumed that the Goldstone-Higgs shift symmetry is exact within the composite sector, and broken only by external gauge and Yukawa couplings.

In our toy models, the effective theory of accidents has a simple ultraviolet completion, in terms of a weakly-coupled and renormalisable theory of an elementary scalar $\phi$ in a large representation of $G$. Notice that, in contrast with models where the Higgs is a NGB, the full symmetry $G$ can be gauged, with no ad-hoc global symmetries assumed: the NGBs are absorbed by the gauge bosons, while the abelian Higgs is a tree-level massless accident. On the other hand, to address the ``big hierarchy problem'' associated with models of elementary scalars, it would be interesting to realise composite accident models, where $\phi$ emerges as a composite multiplet.

%%%%%%%%%%%%%%%%%%%%%%%%%%%%%%%%%%%%%%%%%%%%%%%%%%%%%%%%%%%
\subsection{Accidents in supersymmetry}\label{susyapp}

In supersymmetric models, the accident phenomenon can explain large mass hierarchies: when a symmetry is spontaneously broken at the scale $M$ by some superfield in a large representation, its accident components will remain massless at all orders in perturbation theory, as long as SUSY is unbroken. The super-accident mass scale, which is controlled by SUSY breaking, can be arbitrarily small with respect to $M$. 

One can therefore speculate that super-accidents  might be used to address problems such as doublet-triplet splitting in SUSY grand unification theories (GUTs), see e.g.~\cite{vonGersdorff:2020lll} for a recent discussion and earlier references. This would require a non-trivial generalisation of the toy models above, replacing e.g.~$\SU{2}\times\U{1}$ by some unified gauge group $G$, the residual $\U{1}'$ by the SM gauge symmetry (preserved at some special point in moduli space), and the five-plet $\phi$
with a $G$-multiplet containing an electroweak doublet. The latter should remain accidentally massless to play the role of the SUSY SM Higgs. The accident mechanism would then guarantee the absence of a  $\mu$ term at the GUT symmetry breaking scale $M$, while the other multiplet components do acquire  masses of order $M$, without imposing any ad-hoc cancellations between the superpotential couplings.

Once SUSY is broken, the accident directions are expected to be lifted. Since the moduli space in super-accident models is compact, there is no risk of runaway directions appearing, and the vacuum will be at a stable point in field space. The resulting mass spectrum will depend on the specific mechanism of SUSY-breaking mediation. It would be interesting to study whether the accidental protection mechanism leaves some imprint on the final pattern of soft SUSY-breaking masses. This requires a detailed analysis of concrete example models, which is beyond the scope of the present work.

%%%%%%%%%%%%%%%%%%%%%%%%%%%%%%%%%%%%%%%%%%%%%%%%%%%%%%
\section{Conclusions}

Given a generic renormalisable scalar potential with symmetry $G$ spontaneously broken to $H$, accidents are scalar fields which do not receive a tree-level mass, although they do not belong to the $G/H$ coset. There is no obvious way to infer from symmetry selection rules that the accident masses are suppressed.

We demonstrated that accidents appear in theories with a scalar multiplet in a large representation of $G$. This can be ascribed to the restrictive structure of the most general renormalisable $G$-invariant potential. Already in the minimal models -- the five-plet of $\SU{2}$ and ten-plet of $\SU{3}$ -- the vacuum manifold is non-trivial. It would be valuable to conduct a systematic analysis of the possible field-space geometries leading to accidents.

Accidents possess unsuppressed, tree-level, non-derivative couplings to other scalars. As one moves along the accidental tree-level flat directions, the tree-level mass spectrum of other scalars (and of vectors and fermions which obtain their masses from scalar VEVs) changes, but the sum of the masses squared remains constant.  

When including loop corrections, the flatness of the potential is lifted and an isolated minimum appears, where a non-trivial symmetry $H$ remains unbroken and the number of accidents is enhanced. These models exhibit a one-loop hierarchy of scales, even if all dimensionless couplings are of the same order. It is an open question whether less minimal models exist where an accident mass arises only at two-loop or higher order.

An accidentally light Higgs may help to address the little hierarchy problem.
We built a toy model where an accident plays the role of abelian Higgs with a small VEV, motivating the quest for a less minimal model which could feature a $\SU{2}$ doublet of accidents. We showed that the accident phenomenon persists in supersymmetric theories, and may thus explain large hierarchies among superfield masses without fine-tuning.

The rolling of accidents along their tree-level flat directions provides a playground for natural inflation and/or resonant particle production. Accidentally flat directions lifted by loop corrections may also lead to cosmological first-order phase transitions. Finally, dark-sector accidents are excellent candidates for DM, as they are naturally the lightest states charged under unbroken dark-sector symmetries.

\section*{Acknowledgements}

We thank G.~Burdman and N.~Grimbaum-Yamamoto for useful discussions. MF thanks the IF-USP (Brazil) for hospitality along the development of this work. MF has received support from the European Union Horizon 2020 research and innovation program under the Marie Sk\l odowska-Curie grant agreements No 860881-HIDDeN and No 101086085–ASYMMETRY. TH work is supported by the Excellence of Science (EoS) project No. 30820817 - be.h “The H boson gateway to physics beyond the Standard Model”, and by the IISN convention 4.4503.15.

%%%%%%%%%%%%%%%%%%%%%%%%%%%%%%%%%%%%%%
%%%%%%%%%%%%%%%%%%%%%%%%%%%%%%%%%%%%%%
\bibliographystyle{hieeetr}
\bibliography{accidents.bib}
%%%%%%%%%%%%%%%%%%%%%%%%%%%%%%%%%%%%%%

\end{document}